# Photochromism and influence of point defect charge states on optical absorption in aluminum nitride (AlN)

Ivan Gamov,[*] Carsten Hartmann, Thomas Straubinger, and Matthias Bickermann

*AFFILIATIONS: Leibniz-Institut für Kristallzüchtung, Max-Born-Str. 2, 12489 Berlin, Germany*

\* Author to whom correspondence should be addressed: ivan.gamov@ikz-berlin.de

## ABSTRACT

In this work, we study the absorption properties of AlN in the range of 1.5 – 5.5 eV, as well as the meta-stable change in absorption induced by ultraviolet (UV) irradiation (photochromism), and the restoration of the initial state under the action of irradiation of 2 - 4 eV or elevated temperatures. UV irradiation results in a decrease of the absorption coefficient from 110 cm$^{-1}$ to 55 cm$^{-1}$ at 4.7 eV while in the visible range the absorption coefficient increases from values below 5 cm$^{-1}$ to ~35 cm$^{-1}$. Measurements with two linear polarizations, E∥c and E⊥c, provide determination of several different absorption bands at 2.6, 2.8, 3.4, 4.0, 4.5, and 4.8 eV. The bands at 2.6 eV and 3.4 eV identify the defect levels near to the valence band, while the band peaking at 2.8 eV is related to the conduction band. The photochromism allows controlling the absorption of light in two related spectral ranges because the decrease of UV absorption and increase of the visible absorption are related to switching the charge state of the same defects.

# I. INTRODUCTION

Aluminum nitride (AlN) crystallizes in the wurtzite structure and possesses a direct bandgap of 6.2 eV.[1,2] The large bandgap implies that pure crystals are transparent down to wavelengths of close to 200 nm and have negligible intrinsic conductivity even at high temperatures. Therefore, AlN is a promising material for the fabrication of optoelectronic devices in the deep ultraviolet (UV) range. [3,4] It can also be applied with other III-nitrides in the visible range, and transparency control may be required in both ranges. Suitable AlN bulk crystals are typically grown by physical vapor transport (PVT).[5–7] Optical and electrical properties of the material are determined by intrinsic defects, such as aluminum and nitrogen vacancies ($V_{Al}$,[8–14] $V_N$[15,16]), and extrinsic defects from unintentional background doping by silicon, oxygen, and carbon. These three impurities provide not only isolated substitutional defects (i.e., oxygen and carbon substituting nitrogen, and silicon substituting aluminum: $O_N$, $C_N$, $Si_{Al}$),[17–22] but are also assumed to form complexes, such as $V_{Al}$-$nSi_{Al}$,[23] $V_{Al}$-$nO_N$ (n=1…4 is a number of Si or O atoms),[13,14,16,23–28] $C_N$-$Si_{Al}$,[29] carbon pairs,[30] and tri-carbon complexes.[31] Usually, defects can have several allowed charge states in the bandgap,[32,33] including DX-states,[34–38] resulting in the polymorphism of defects properties. The current charge states of defects are given by the Fermi level which should be known first for a successful characterization of AlN crystals.

In this work, we present a detailed investigation of photochromism in AlN that allows us to determine the Fermi level and the active defect levels in the bandgap. Photochromism is a reversible, yet room-temperature meta-stable change of the optical properties upon irradiation with light (here, UV irradiation). Color centers ("F centers") in alkali halides are the classical example of photocromism.[39] Furthermore, photochromic effects are found in a large variety of other crystals such as $CaF_2$, GaN, $SrTiO_3$, and synthetic diamonds.[40–46] The key feature of photochromism is the stability of the induced properties even after the radiation ended, which can be caused by two different reasons: (i) physical change of the defects due to irradiation, i.e. the light-induced dissociation of defect complexes or strong local lattice relaxation (including light-induced re-bonding) as observed in GaN[40,41] and fluorides,[42,43] and (ii) change of the defect population, e.g. *the band-mediated charge transfer* between defects as proposed for synthetic diamonds[44] or SrTiO$_3$[45,46]. The first study of photochromism in AlN (as ceramic) dates back to 1992,[47] in which the authors suggested that the creation of UV-induced meta-stable states for $V_{Al}$-$O_N$ would give rise to a change in absorption and emission. However, in the present work we show that photochromism in AlN originates from *the band-mediated charge transfer* which results in a changed (meta-)stable population of ordinary point defects.

Based on the evidenced model, we evaluate the Fermi-level positions and the energy levels of the involved defects comparing AlN with different chemical concentrations of silicon, oxygen, and carbon. Furthermore, the observed effects are significant and should be accounted for in the development of optical devices.

## II. EXPERIMENTAL

The AlN single crystal was grown by physical vapor transport (PVT) via spontaneous nucleation at the Leibniz-Institut für Kristallzüchtung (IKZ).[21,48] The crystal was cut parallel to the m-plane (perpendicular to the growth plane), out from the central area of the bulk crystal, and then chemo-mechanically polished on both sides. The sample is shown in Fig. 1, three areas of the interest (B1 – B3) are marked. Two main factors contribute to the concentrations of impurities: first, there are two different growth facets[19] (areas B2 and B3 were grown on -c [000-1] and area B1 on m [10-10] facets) and second, there is a transient effect during growth: the oxygen supply in the crucible and the incorporation of oxygen into the crystal decreases with growth time (from B3 to B2). The concentrations of the main impurities, i.e., oxygen, carbon, and silicon, in three different sample areas B1-B3 were determined by secondary ion mass spectrometry (SIMS) at RTG Mikroanalyse GmbH Berlin. Table I summarizes the SIMS data for these areas in the regions of interest, together with the nominal donor-to-acceptor impurity ratio R = ([Si]+[O])/[C].

The optical absorption was investigated in areas B1 – B3 shown in Fig. 1 bordered by a rectangular aperture ($0.5 \times 3$ mm$^2$) and held precisely during all manipulations. In these areas, the respective SIMS analysis within $0.2 \times 0.2$ mm$^2$ regions is also performed. The optical transmission spectra are recorded at room temperature in the UV-visible range (225–800 nm) using a double beam, double monochromator spectrophotometer (PerkinElmer LAMBDA 1050). Measurements with polarized light are recorded by two Glan-Thompson polarizers inserted in the sample and reference beam paths; the polarizers are efficient for wavelengths longer than 225 nm. Two linear polarizations, E⊥c and E∥c, are investigated (the orientation of the light electric field vector $E$ respective to the c-axis is shown in Fig. 1). The resulting absorption coefficients $\alpha$ are corrected for reflection (the Fresnel correction). The thickness of the investigated AlN sample was about 500 μm; to measure absorption coefficients exceeding 140 cm⁻¹ in the B1 area, the sample was subsequently thinned down to 110 μm.

Two parameters derived from the transmission spectra (optical transmission $T = I/I_0$ at a given wavelength $\lambda$, $I_0$ and $I$ as incoming and transmitted light intensities) are used in the current work. First, the wavelength-dependent relative change in transmission ($CIT$) caused by UV irradiation or annealing is determined using the formula: $CIT = (T_a-T_b)/(T_a+T_b)$, where $T_a$ and $T_b$ are the transmission after and before the influence, respectively.[1] Likewise the second parameter, the wavelength-dependent degree of polarization $DOP = (T_{/\!/} - T_{\perp})/(T_{/\!/} + T_{\perp})$ is used to compare the transmission for two corresponding polarizations ($T_{/\!/}$ and $T_{\perp}$ for E∥c and E⊥c, respectively).

The UV irradiation is performed using either the diffused light of a Hg-lamp (with the band-pass interference filter selecting the Hg-line at 254 nm / 4.88 eV) or the line at 266 nm (4.66 eV) of a Nd:YAG continuous wave laser (FQCW266, CryLaS GmbH). The Hg-lamp nominal electric power of 6 W (Heraeus TUV6WE) provides less optical power density than the laser beam which has a nominal optical power of 55 mW, focused in the aperture areas. In contrast, the spectrometer light source does not change the sample state even after multiple measurements. The visible light irradiation is performed with LEDs emitting around 2.25 eV (550 nm, nominal optical power 3W) or 3.2 eV (385 nm, 50 mW) and with defocused lasers emitting at 2.41 eV, 2.54 eV, 2.81 eV, or 3.81 eV (514, 488, 442, 325 nm, respectively; nominal optical power 50–100 mW).

The annealing is performed in several steps with a sequential increase in time and/or temperature, as illustrated by the time-temperature dependence (insert in Fig. 3 (b)). The sample which is glued on one side to the optical aperture (aluminum plate 40x50x4 mm$^3$) is brought in contact on the other side with a massive heating plate (covered with Al foil) with an internal temperature controller. No additional UV irradiation is performed between the steps. All optical measurements are carried out after cooling down to room temperature.

---

[1] In particular, in Fig. 5 (c, d), the *CIT* corresponds to the change in transmission 5 minutes after the UV laser irradiation ($T_a$) compared to that before the irradiation ($T_b$) and in Fig. 7, the stabilized transmission measured 1 hour after the UV irradiation ($T_b$) is compared to the transmission after two different annealing steps ($T_a$).

## III. RESULTS

The room temperature absorption spectra are shown in Fig. 2 for the three investigated areas B1, B2, and B3. The measurement of the m-cut sample for E⊥c polarization corresponds to measurements of c-plane AlN wafers reported previously.[6,15,49,50] Several unresolved absorption bands in the range below 4.0 eV are observed for areas B1 and B3, while in area B2 no absorption bands are found ($\alpha$<5 cm$^{-1}$), so area B2 appears colorless to the eye. The band peaking at 4.7 eV, usually referred to as "carbon related", was observed in areas B1 and B2; this was discussed previous studies also from our group.[15,29–31]. For these two areas, a clear correlation between the carbon concentration [C] and $\alpha$ at 4.7 eV is observed: for B1 [C]=5.7×10$^{18}$ cm$^{-3}$, $\alpha$~1100 cm$^{-1}$ and for B2 [C]=1.8×10$^{18}$ cm$^{-3}$, $\alpha$=115 cm$^{-1}$; the absorption band spans exactly the same energy range. However, in area B3 there is no absorption band at 4.7 eV although this area contains the same carbon concentration as in B2.

We note that pronounced photochromism appears in area B2, see Fig. 3(a). This area has a lower carbon concentration than that of area B1, a lower oxygen concentration than that of B3, and corresponds to the intermediate donor-to-acceptor ratio R=2. Although this study focuses on only one AlN crystal, similar properties were found in several other AlN crystals with similar values of R. In Fig. 3 (a), the absorption at around 4.7 eV is significantly reduced after UV irradiation, and the absorption coefficient $\alpha$ decreases from 115 to 60 cm$^{-1}$. Simultaneously, in the spectral range below 4.2 eV and from 5.0 to 5.4 eV, the absorption increases.

The UV irradiation process is described in detail in the Experimental section. The spectral curves taken prior to the irradiation (room-temperature equilibrium population of defects, the black curve) and after long irradiation (2 minutes) under intense laser light at 266 nm (4.66 eV, the green curve) correspond to the maximum and minimum $\alpha$ at 4.7 eV observed in the area, respectively. The intermediate curves illustrate a similar but weaker influence of the Hg-lamp irradiation at 254 nm (4.88 eV) with lower power density. The dose of the UV irradiation was accumulated in the AlN crystal (from 5 to 80 minutes, the duration of the irradiation increased with each step and there was no restoration of the initial state between the steps). With the accumulation of the UV irradiation dose, the change in transmittance (or absorption) was monotonically enhanced in the whole observed spectral range and gradually approached saturation.

Remarkably, the changes are meta-stable at room temperature, as the absorption coefficients change only a few percent per day. However, fast and complete restoration of the initial absorption spectrum can be accomplished in two ways: with annealing, and with intense irradiation in the range of $2-4$ eV. Fig. 3 (b) shows the absorption spectra after the annealing at different temperatures (duration and temperatures are shown in the insert of Fig. 3). Measurements after each step p0 to p8 are shown as absorption spectra of individual color. The red spectral curve with the highest $\alpha$ at 2.8 eV corresponds to the measurement immediately after the UV irradiation, and the light-brown area bordering by this curve indicates how the absorption changes in the dark during the first 5 minutes after the UV irradiation (p0 – p1). This change, likely due to homogenization and equilibration after the UV irradiation, is not discussed in this paper. After this rapid effect, further measurements at room temperature do not reveal significant changes; the corresponding minimal difference in the absorption spectra taken 5 (p1) and  60 (p2) minutes after the UV irradiation at room temperature is shown by the dark-brown area.

The changes upon further annealing strongly vary for different spectral ranges. The band at 2.8 eV disappears first (the pink areas in Fig. 3(b) between p3 and p5, annealing at 450 K), the peak of absorption shifts to 2.6 eV. The decrease of the induced absorption at 4.0 eV begins only at 600 K (the blue area between p7 and p8, 600 K). Only at higher temperatures, all photochromic changes are clearly restored to the initial state. In contrast, restoration using intense irradiation in the range from 2–4 eV (see details in the Experimental section) seems to proceed with conditionally proportional changes in transmission at different spectral ranges until it reaches complete restoration (the spectra are not different from the spectra in the initial state). Also, the effect is virtually independent on the actual wavelength of irradiation in this wavelength range.

Additional polarization measurements are required to reliably identify the defects involved in the photochromism. In Fig. 4, the absorption spectra for E⊥c (solid lines) and for E∥c (dashed lines) are shown for areas B1 – B3 before (a, c, e) and after (c, d, f) UV irradiation. The UV-induced changes in area B1 are negligible and unstable, i.e., they disappear within ~30 minutes after the irradiation ended. Apparently, the charge state of the corresponding defect cannot be significantly changed by UV irradiation. One explanation might be that due to the low penetration depth (below 10 μm) for 4.7eV radiation at high $\alpha$~1000 cm$^{-1}$, the defects in the bulk volume remain largely unaffected: actually, gradual absorption changes appeared in a similar sample after X-ray irradiation that were meta-stable (did not change after 2 weeks under ambient conditions).[31] The changes in area B3 are comparable with the most stable and strong effects in area B2. Additionally, in Fig. 4 (c, d), the DOP is calculated for area B3 before and after the UV irradiation. The bands peaking at 2.8, 3.2, 3.4, and 4.0 eV are marked for different areas in the figure and are listed in Table II with the other bands.

In Fig. 5, further analysis of the polarized absorption is illustrated to provide a clear comparison of the experimental data. The DOP is shown in Fig. 5 (a) for area B1 before and after the UV irradiation, and for B2 before, 5 min after UV irradiation (p1 in Fig. 3 (b)), and after the last step of annealing (p8 in Fig. 3(b)) at 600 K. The spectral curves for the two areas are very similar and the incomplete thermally induced restoration does not change the shape of the DOP curve significantly. Unfortunately, the absorption spectra above 4.3 eV for area B1 could not be measured even after the thinning down of the sample; thus, the DOP curves are also not available. That's why we included DOP data from another AlN crystal grown in our laboratory with [C]=5.8×$10^{18}$ cm$^{-3}$, [O]= 2.4×$10^{18}$ cm$^{-3}$, and [Si]= 3×$10^{16}$ cm$^{-3}$, $\alpha \approx 1600$ cm$^{-1}$ at 4.7 eV,[30] thus similar to area B1. Positions and magnitudes of the minimum (at 4.3-4.4 eV) and the maximum (at 4.9 eV) coincide for both DOP curves, despite of the clearly different impurity concentrations in the samples. In accordance with the highest peaks from the Gaussian approximation in Fig. 5 (b), the low-energy and the high-energy ranges seem to center at around 4.5 eV and 4.8 eV, respectively. We attribute these components to different defects and discuss separately. The parameters of the peaks are given in the insert. Although at least three peaks are required for adequate fitting of the E⊥c and E∥c polarized spectra, we avoid further separation of the peaks. The 4.5 eV peak prevails for E∥c polarization when the 4.8 eV peak is depolarized or demonstrates a weak E⊥c polarization (DOP can be distort by overlapping with E∥c-polarized bands). Finally, Fig. 5 (c, d) visualizes the difference of CIT for areas B2 and B3. The strong "darkening" peaking at 5.2 eV in area B2 is listed as an induced absorption peak in Table II.

A closer look at the thermal restoration (Fig. 6) uncovers specifics of the restoration processes that are similar in samples B2 and B3. For sample B2, CIT is calculated after the two annealing steps at 450 K and at 600 K (p5 and p8 in Fig. 3(b), respectively) with respect to the transmission spectra recorded 60 minutes after UV irradiation at room temperature (p2 in Fig. 3(b)). Notably, for p5 the absorption bands at 4.0 eV and 4.5 eV continue to increase and decrease, respectively, even 120 minutes after the UV irradiation. In contrast, absorption at 2.5 – 3.5 eV starts to restore (decrease) immediately at 400 K, and only at 600 K all UV-induced peaks are clearly being restored. The weakly evident induced absorption band at 2.6 eV is more stable than the band at 2.8 eV and can be observed after the annealing at 450 K, as shown in Fig. 3 (b). The band at 2.6 eV (peaking in Fig. 3 (b) p8) is included in Table II separately from the band at 2.8 eV.

In Fig. 6, the UV-induced CIT for area B3 (E⊥c) is taken from Fig. 5 (d) at a modified scale ×0.5 (the blue curve), to show that this curve precisely coincides with the one for area B2 at the same polarization (red solid curve).

## IV. DISCUSSION

Based on the results presented above, the origins of the observed photochromism in sample area B2 are proposed. We note that our results are in perfect agreement with the model of the band-mediated charge transfer between defects. Different defects in the initial states participate in the original UV absorption but are transparent for the visible light. Then, after optical ionization, they are involved in the charge exchange, while maintaining general electric neutrality. Some defects (we call them D1-like) play a role of electron donors, and others accept electrons (D2-like). A direct charge transfer (the donor-acceptor pair (DAP) transition) does not lead to the meta-stable carrier redistribution due to the equal probabilities of exciting and relaxation transitions as the optical transition probability is independent on the transition direction according to the Fermi's golden rule. Thus, in the photochromic AlN crystal the spatial distance between defects is expected to be larger than the DAP distance, but small enough to facilitate charge transfer. In this case, the non-equilibrium meta-stable population is provided by transfer of photo-induced carriers via the conduction and valence bands (CB and VB). By capturing charge carriers, defects change their charge states and subsequently are able to absorb light in other spectral ranges. Initiating the reverse transfer leading to restoration of the initial state requires additional energy to inject the carriers back into CB or VB. We propose $C_N$, $C_{Al}$, $C_N$-$C_{Al}$, and $O_N$ defects as possible candidates for most of the induced absorption changes.

*Experimental evidence of the change of the defect charge state*

The absorption spectra provide reasonable evidence that the absorption at $2.0 - 3.8$ eV in areas B1 and B2 is of the same origin, as the spectra have the same peak energy and a very similar DOP (see Figs. 4 (a, b, d) and 5 (a)). In area B2, the DOP decreases due to the annealing and the induced absorption also decreases, however, the curves preserve their characteristic two-step form, and the spectrum remains very close to the DOP of the sample area B1, which mostly does not depend on annealing or radiation. The similarity of DOP implies the same defects at comparable concentrations exist in B1 and in B2 after the UV irradiation. This conclusion is more reliable for the strictly polarized bands and less for depolarized absorption bands. Since in B1 the defects are stable and exist independently on visible and UV irradiation or thermal influences, they cannot be treated as light-induced (caused by radiation). Consequently, also the meta-stable induced absorption in area B2 cannot be attributed to light-induced defect generation, radiation damage, or dissociation of defect complexes.

We have shown that at least a fraction of inactive defects in area B2 is activated by UV irradiation and then deactivated thermally or by $2 - 4$ eV irradiation. Such "(de-)activation" typically corresponds to a switching between two charge states of the defect. Thus, in B1 and B2 areas, the defects are originally in different charge states, i.e., initially the sample areas have different positions of the Fermi level determined by the proportion of donor-like and acceptor-like point defects in these areas.

*The band-mediated charge transfer model*

We propose the model of band-mediated charge transfer between defects D1 and D2, see Fig. 7. In the specific example shown in Fig. 7 (a), an electron which initially belongs to defect D1 is being excited by $E_{UV}$ (D1) and then transferred via the conduction band to another remote unoccupied defect (D2).[2] Analogously, the transfer of a hole is possible via the valence band (excitation $E_{UV}$(D2)) and leads to the identical result. The possibility of the transfer implies the absence of immediate recombination of the UV-excited carriers to their initial states (on D1 or D2). On the contrary, the carriers, driven by Coulomb interactions or due to Brownian motion, reach the remote defect that can capture the electron (or hole) effectively.

We show how the change of the D1 and D2 population influences the optical properties. In thermodynamic equilibrium, the Fermi level ($E_F$) determines the initial charge states for defects D1 and D2 as $i$ and $j$, respectively. We assume the range of the charge states of D1 and D2 spans from ($i$-$1$) to ($i$+$2$) and from ($j$-$2$) to ($j$+$1$), respectively, thus, in general four absorption bands are possible in the equilibrium before the UV irradiation, these transitions are shown by arrows labeled $E_{UV}$ and $E_o$ in Fig. 7 (a, b). The number of defects, D1 and D2, in the initial charge states and the related UV absorption both decrease due to the UV irradiation ($E_{UV} \approx 4.7$eV). In contrast, the UV irradiation leads to an increase of population for the nonequilibrium charge states $i$+1 and $j$-1 and the related absorption in the visible range ($E_{Vis} \approx 2 - 4$ eV) and UV range ($E_{ind} > E_{UV}$). Thus, another four absorption transitions labelled $E_{Vis}$ and $E_{ind}$ can be active for the non-equilibrium population.

The simple inequality $E_{ind} > E_{UV}$ holds for the case of normal point defects with the sequential arrangement of the transition levels (in contrast e.g. to "negative-U defects" such as DX-like centers). In addition, some relations for energies, $E_{UV}$ and $E_{Vis}$, can be given. If $E_{UV} =$ $E_{UV}^{ZPL} + E_{UV}^{FCS}$, ($E_{UV}^{ZPL}$ - zero phonon energy), the energy $E_{Vis} = E_{Vis}^{ZPL} + E_{Vis}^{FCS}$ can be estimated from the bandgap as:

$$E_g = E_{Vis}^{ZPL} + E_{UV}^{ZPL} \approx 6.2 \text{ eV} \qquad (1)$$

where $E_{Vis}$ is a complimentary transition for $E_{UV}$ as shown in Fig. 7. The included Franck-Condon shift energy ($E^{FCS}$) is a part of the energy dissipated by vibrations, i.e., the difference between the detected absorption peak maximum energy and the defect's charge carrier energy in respect to the CB or VB (i.e., its "position in the band gap"), which for the particular optical transition is equal to the zero phonon energy $E^{ZPL}$ (not the difference between the emission and the absorption maxima). The typical values of $E^{FCS}$ are discussed in the next section.

---

[2] I.e., D2 is initially occupied by a hole

The diagrams in Fig. 7 (c, d) illustrate the restoration process of the AlN crystal with 2–4 eV irradiation or due to annealing at elevated temperatures. Transitions labeled $E_{Vis}$ initiate the reversed carrier transfer and return the system to its initial state. Optical activation of these transitions in the experiment is possible by excitation in the range of 2 – 4 eV. If the related defect levels are not very deep, the defect can be also ionized thermally (600 K successfully reverts all induced absorption changes). The processes of restoration can promote emission/luminescence (arrows $E_{Res}(D1)$ and $E_{Res}(D2)$) at energies even higher than the optical excitation energy. We suggest that such emission was observed by Okada et al. after X-Ray/UV irradiation of AlN ceramic samples[51–53] or by Spiridonov et al. after UV irradiation of high-oxygen AlN.[54]

Below we discuss the absorption bands appearing after the UV irradiation in area B2. The defect levels and the optical transitions are visualized in Fig. 8. The blue arrows (4.5 and 4.8 eV) illustrate the processes of the original absorption in the sample area before the UV irradiation, and the orange arrows (2.6, 2.8, 3.4, and 4.0 eV) correspond to the UV induced absorption bands. We also schematically show there the charge transfer via CB and VB accompanying the absorption changes at different temperatures and candidates for the absorption bands are proposed.

*Considerations on the identification of the defect levels*

The thermal equilibrium absorption involves the charge states of defects determined by the Fermi level. Sample areas B1 – B3 differ in their concentrations of impurities and point defects that determine the Fermi level position. The donor-to-acceptor ratio R is calculated using chemical analysis assuming all oxygen, silicon, and carbon atoms form effective donors ($O_N$, $Si_{Al}$) and acceptors ($C_N$).[23,24,55] However, for N-rich growth conditions of AlN with the significant influence of acceptors as in areas B1 (R=0.7) and, possibly, B2 (R=2.0), also $C_{Al}$ might manifest as a donor-like compensating defect; the formation energy of $C_{Al}$ is among the lowest for carbon defects discussed in literature.[55–58] Furthermore, $C_{Al}$ is suggested as a donor defect found in AlN by means of optical and electron-resonance methods.[59,60] Also, the $V_N$ defect is expected as a pronounced compensating donor defect in p-type material.[15,16,61] Carbon pairs and tri-carbon defects exist in the bulk AlN as was found experimentally[30,31]. Similarly, relatively high oxygen concentration in areas B2 as well as B3 (R=3.7) can lead to significant formation of $V_{Al}$ and ($V_{Al}$-$nO_N$) compensating centers, as their formation energies can be comparable to $O_N$ in N-rich growth of AlN with domination of donors.[23,24,58,62] Nevertheless, we suggest that there is a direct correlation between the positions of the Fermi level in different regions with the ratio R.

We assume sample area B2 is perfectly compensated and has the Fermi level most close to the middle of the bandgap, while in areas B1 and B3 their equilibrium Fermi-levels are shifted respectively down and up from the position of B2, i.e., the empty defect states near the VB and the occupied states near the CB appear. These states lead to the visible range absorption in areas B1 (2.6 eV and 3.4 eV) and B3 (2.8 eV). The UV excitation in area B2 results in a changed population of the defect levels that activates both the B1-like and the B3-like absorption features in the area B2. Fermi level shift and the associated change in optical due to thermal diffusion of beryllium in AlN were reported by Soltamov et al.[49,63]

Hence, the photochromic point defects play a role of electron and hole traps, and the corresponding trap energy ("depth") $E^{ZPL}$ can be evaluated from the annealing temperature or from the energy of optical absorption peak energy ($E_A = E^{ZPL} + E^{FCS}$). Summarizing the theoretical estimations of $E^{FCS}$ energy of different defects,[24,29,50,55,62] and the difference between the emission peak energy and $E^{ZPL}$ which has the same vibrational nature as $E^{FCS}$,[58] the values of $E^{FCS}$ can be in the range of 0.3 - 1.0 eV (typically 0.5 - 0.8 eV). The highest value is attributed to $V_N$ vacancies[58] since the heavy surrounding Al atoms vibrate at very low frequency. In case of DX-states, the energy difference between the thermal activation depth and the absorption peak can be significantly higher;[37,38,64–66] probably up to 1.8 eV. In the following sub-sections, we detail and summarize the data for each of the absorption bands in the B2 area of the AlN sample.

### *The photochromic induced absorption bands 2.6, 2.8, 3.4, 4.0, and 5.2 eV in area B2*

The 2.8 eV induced absorption band reverts back to the initial state at lower temperatures (400 K) compared to the other bands in area B2 (p3 in Fig. 3 (b)). DOP of the 2.8 eV band can be estimated as -0.1 before and after the amplification by the UV irradiation in area B3 (Fig. 4 (e, f)), where the band prevails in its spectral range. The related optical transition is assumed to take place from electron-occupied defect states in the upper half of the bandgap, as the high [O] (R=3.7) as expected, results in a high position of the Fermi level due to prevalence of donor-like $O_N$ defects (see e.g. [62]). The band has been earlier reported to be polarization-dependent in oxygen-dominated samples and its intensity correlates with high oxygen concentrations.[6,26,67,68] From the absorption peak energy and the temperature of the thermally-induced restoration, the position below CB is estimated to be 1.0 – 2.5 eV and 1.0 – 1.2 eV, respectively. Thus, both thermal and optical activation of the state can be associated with the same defect level, only if this value is roughly close to 1 eV. The $O_N$ defect in the DX-configuration corresponds well and can explain the large $E^{FCS} \approx 1.8$ eV. However, the "same defect" assumption requires further experimental evidence.

The absorption band at 4.0 eV, only evident in area B2, is the most thermally stable induced band in the 2–4 eV range. Even at temperatures of 600 K (see p8 in Fig. 3 (b)), the direct thermal activation from the defect level seems unlikely. The related defect behaves as a very deep trap capturing charge carriers released from other traps, e.g., electrons at medium temperatures (300 - 450 K), when the absorption still increases, or holes at higher temperatures, when the induced absorption decreases and finally disappears. The DOP of the band is positive (stronger for E∥c) but it is also influenced by the overlapping band at 3.4 eV (E∥c), as shown in Fig. 4 (d). A defect with suitable properties would be $C_{Al}$ (i.e., in DX2 ground state configuration, $E^{ZPL} \approx 2.7$ eV)[38] or the carbon pair defect (absorption between 3.8 – 5.1 eV is expected)[30]. Different charge states of $C_{Al}$ are expected in the range 2 – 2.7 eV below CB,[38,55,57] which would fit well to the 4.0 eV induced band and also to the 4.5 eV band generally observed in carbon-dominating samples ($E_{Vis}$(D2) and $E_{UV}$(D2) transitions in Fig. 7, respectively), or both, since the relaxation to the DX2 state gives a large additional energy shift.[38]

The band at 2.6 eV becomes prominent in area B2 of the AlN crystal after partial annealing (Fig. 3 (b)) and is clearly seen in area B1, while it does not arise at all in area B3. Its restoration starts above 550 K (p5), thus a direct thermal activation is probable, considering that the $E^{ZPL}$ is in the range from 1.4 to 1.9 eV above the valence band, as we expect in acceptor-doped area B1 (R=0.7). It could be also well correlated with the concentration of carbon. The $C_N^{1-/0}$ transition level fits, as its energy above VB is reported to be around 1.8–1.9 eV.[15,55,57] This level is also discussed in relation to the original absorption at 4.8 eV.

The 3.4 eV band shows up in B2 area as an induced band and also is prominent in B1 area. Assuming that this band should be very broad, we also associate it with high DOP values (~0.35) in both sample areas. We distinguish it from the 3.2 eV band in B3 area, since although the latter also shows a pronounced positive DOP (~0.3) they have different shapes. The 3.4 eV absorption band could be associated with the transitions from VB to different charge states of $V_{Al}$-$nO_N$ complexes, as calculated and observed in other studies,[24,32,62] or with excitations of electrons at $V_N$ defects which should demonstrate a large value of $E^{FCS}$ (~1 eV difference between the photoluminescence peak and $E^{ZPL}$ is reported)[58]. Abundance and configuration of $V_{Al}$-$nO_N$ defects might depend on the oxygen concentration, but we do not have enough evidence to propose a particular configuration.

Finally, the 5.2 eV band could be an $E_{ind}$-type transition (Fig. 7), i.e., the secondary ionization of a defect with an original absorption of 4.5 eV or 4.8 eV.

### *The photochromic original absorption bands around 4.7 eV in area B2*

The peak at 4.5 eV dominates for E∥c polarization while the peak at 4.8 eV shows the opposite effect. The energy of the complementary transition $E_{Vis}$ for the 4.8 eV peak can be estimated as $E_{Vis} \approx$ 2.6 eV (using eq. (1) and $E_{UV}^{FCS} \approx E_{Vis}^{FCS} \approx$ 0.6 eV). This indicates that both 2.6 eV and 4.8 eV bands can be attributed to the same transition level as mentioned previously, $C_N^{1-/0}$. Likewise, the 4.4-4.5 eV band and the induced 3.4 eV band have high DOP in the range of 0.35–0.5. Here, $E_{UV}^{FCS} \approx E_{Vis}^{FCS} \approx$ 0.8-0.9 eV seems high but it is still a reasonable value for AlN (as shown above). As the absorption 3.4 eV appears in B1 (R=0.7, acceptors dominate), this transition likely takes place in the lower part of the bandgap (transition type $E_{Vis}$(D1) in Fig. 7).

## V. CONCLUSION

The investigation of photochromism in AlN provided important information about the defect levels and allows a comparison of the Fermi level positions in different sample areas. In particular, we found pairs of complementary absorption bands 2.6 eV - 4.8 eV and 3.4 - 4.5 eV. The 2.6 and 3.4 eV bands dominate for E∥c polarization and can serve as an identifier of the Fermi level near the VB. The contrast, the 2.8 eV band has a different polarization and indicates that the Fermi level is closer to the conduction band. Further investigation of the activation and deactivation of absorption bands might be possible using controlled co-doping, e.g., with beryllium.

From a practical point of view, photochromism can lead to significant "UV bleaching" (decrease of the UV absorption coefficient under UV irradiation) in AlN. This is relevant for visible range and UV emitting devices on AlN substrates. Furthermore, the spectrometer-measured absorption coefficient values will differ from practical values depending on the irradiation power. In the investigated AlN crystal, the absorption coefficient $\alpha$ at 4.7 eV decreases from ~110 to ~55 cm$^{-1}$ in area B2, while in area B1 such changes were not observed due to $\alpha > 1000$ cm$^{-1}$. The decrease in 4-5 eV UV absorption is, in contrast to optical absorption saturation, accompanied by appearance of an induced absorption in the range of 2–4 eV. The induced changes are meta-stable at room temperatures, but are completely reversible at temperatures above 600 K or by adequate irradiation in the range of 2 – 4 eV. Hence, the optical transmission of AlN in the visible and UV ranges depend on each other and one can be controlled by other.

We suggest that the photochromism is significant particularly when the compensation, i.e. the balance between the specific defects, is nearly ideal such as in area B2 of the investigated AlN sample. Such conditions might provide high UV transparency in the AlN single crystals even without extreme purity. The role of point defects in AlN could be further clarified by additional inspection of the emission stimulated by the restoration of the initial state after the UV irradiation, which was however beyond the scope of this paper.

ACKNOWLEDGMENT

The authors thankfully acknowledge contributions of Dr. Klaus Irmscher, Dr. Yuri Kogut (TU Clausthal), and Dr. Mariia Anikeeva to this work. The research was supported by the German Research Foundation (DFG) under project No. BI 781/11-1.

DATA AVAILABILITY STATEMENT

The data used in the study as well as further data that support the findings of this study are available from the author (I.G.) upon reasonable request.

FIGURES

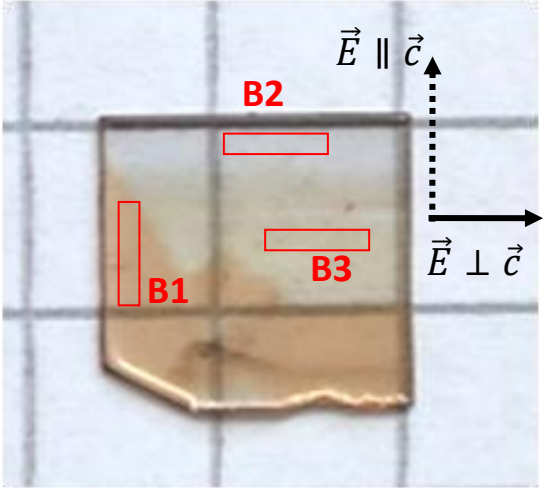

Fig. 1. AlN single crystal cut in m-plane (nearly square size, edge length ~7 mm). Areas B1 – B3 are positions of optical measurements that also contain the locations of the respective SIMS probes. The directions of the electric field vector of the incident light (the light is directed towards the surface normal) are shown regarding the crystal c-axis for two linear polarizations, E∥c and E⊥c.

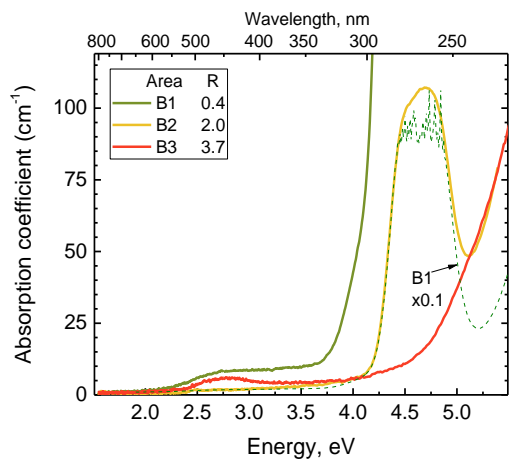

Fig. 2. UV-Vis absorption spectra for E⊥c polarization for sample areas B1–B3 characterized by different values of R = ([Si]+[O])/[C]). The absorption spectrum of area B1 additionally measured after thinning down the sample is presented on reduced scale (×0.1, dotted curve).

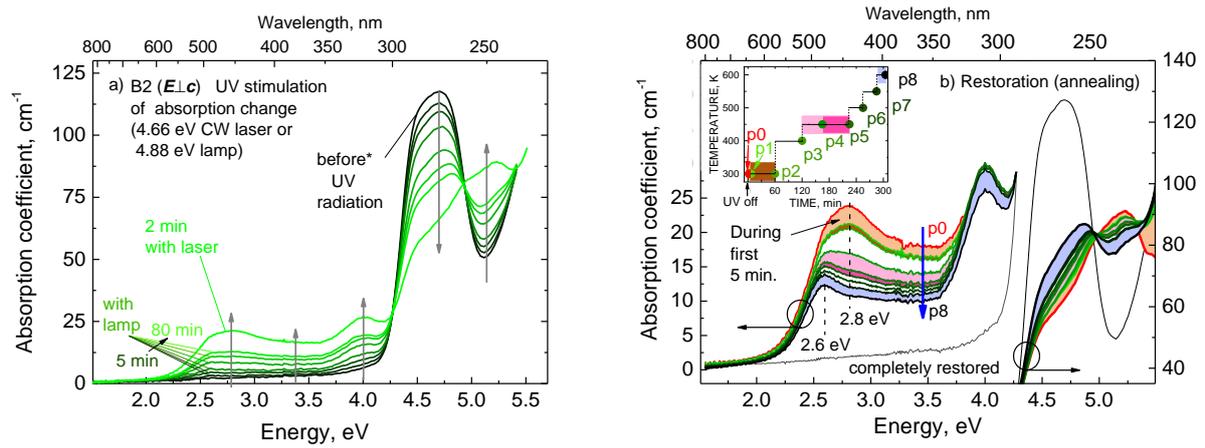

Fig. 3. (a) Stimulation of the absorption change (photochromism) in area B2 by UV irradiation (2 min of the CW laser at 4.66 eV or 5-80 min of 4.88 eV Hg-lamp) and (b) restoration of the initial state by thermal annealing. The time and temperatures of the annealing steps are shown in the insert. Spectra p0 – p8 were measured at room temperature after completion of the respective successive annealing step, respectively. All spectra are measured for E⊥c polarization.

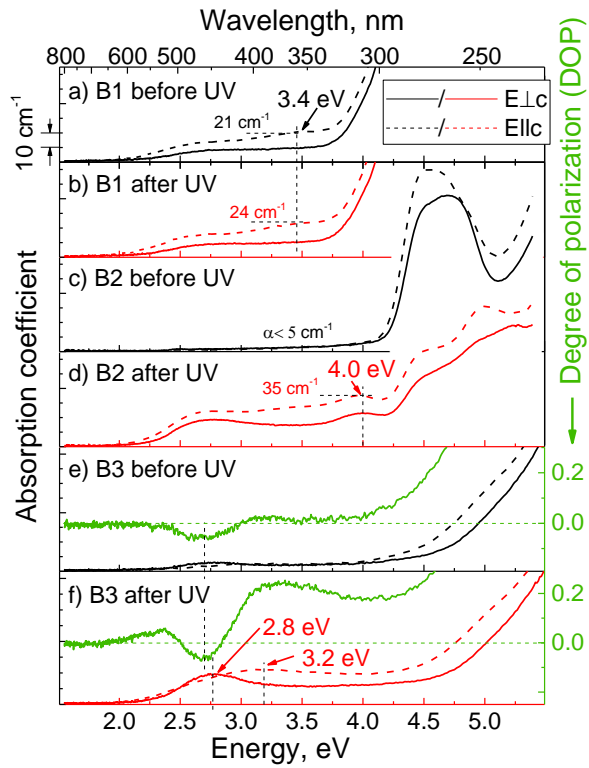

Fig. 4. The absorption spectra before (a, c, e) and after (b, d, f) the UV irradiation for areas B1- B3 for two polarizations of light: E⊥c (solid lines) and E∥c (dashed lines). The scale mark "10 cm⁻¹" is relevant for all absorption spectra. The most pronounced bands in the range 2 – 4 eV are marked. The degree of polarization (DOP) values for area B3 before and after UV irradiation (green lines) correspond to the right x-axis.

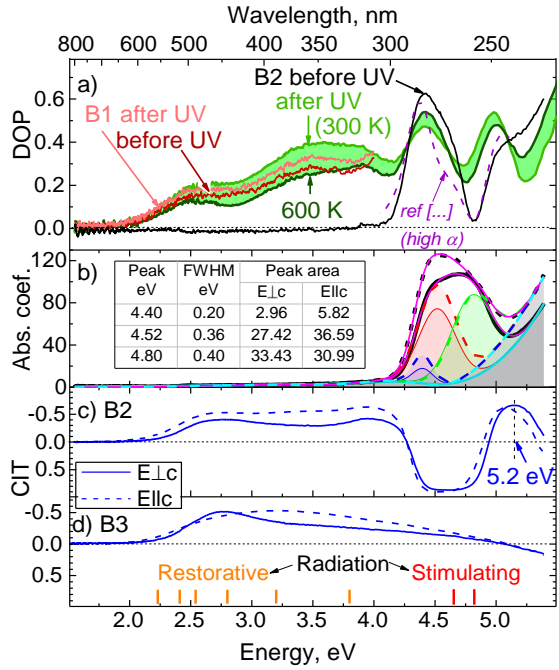

| Peak eV | FWHM eV | Peak area E⊥c | E∥c |
|---|---|---|---|
| 4.40 | 0.20 | 2.96 | 5.82 |
| 4.52 | 0.36 | 27.42 | 36.59 |
| 4.80 | 0.40 | 33.43 | 30.99 |

Fig. 5. (a) The degree of polarization (DOP) for areas B1 and B2 before and after UV radiation. For area B2, the change of DOP after the last annealing step (p8) is shown as green shaded area. In the range above 4 eV where data for B1 are not available, DOP for another AlN crystal with similar properties (see [30]) is reproduced (purple dotted line). (b) Multi-component Gaussian fitting of the absorption band at 4.7 eV for E∥c and E⊥c polarizations. The peak parameters are listed in the insert. (c, d) The change in transmission (CIT) due to UV radiation for areas B2 and B3 and polarizations E⊥c (solid lines) and E∥c (dashed lines). The induced absorption band at 5.2 eV in area B2 is marked. The energies for the restorative and the stimulating irradiation are represented by the orange and red vertical lines on the bottom axis.

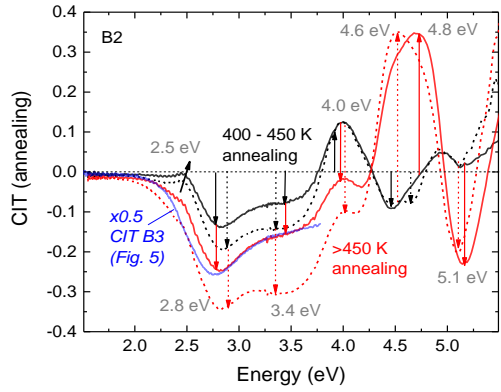

Fig. 6. The change in transmission (CIT) for polarizations E⊥c (solid lines) and E∥c (dashed lines) after annealing step p5 (black lines) and p8 (red lines), i.e. 450 K and 600 K. The CIT is referenced to the transmission 60 minutes after laser UV irradiation (p2). The blue curve duplicates the CIT of sample B3 from Fig. 4 (d) for comparison.

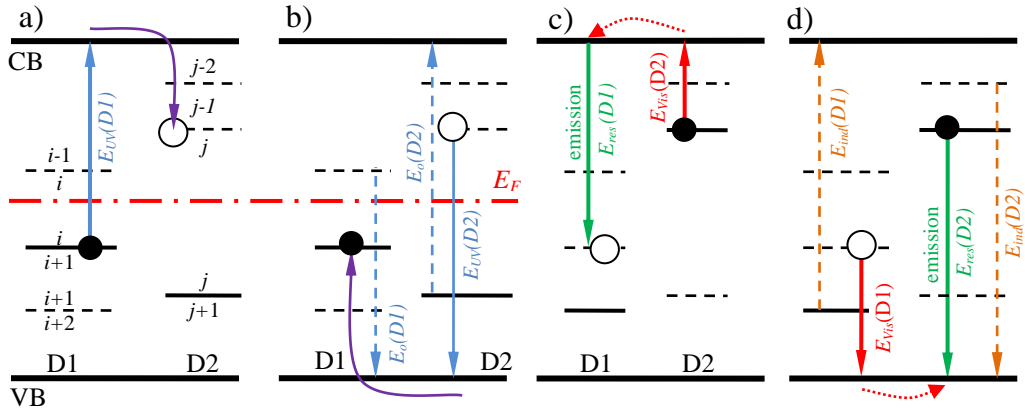

Fig. 7. Energy diagrams for the CB-mediated (a) and VB-mediated (b) charge transfer caused by the UV radiation (with $E_{UV}$ energy). In general, four different absorption processes could appear for D1 and D2 (blue arrows $E_{UV}$ and $E_o$) in their initial charge states $i$ and $j$. The equilibrium defect population is shown schematically by $E_F$ (the states below $E_F$ are occupied by electrons, and above $E_F$ are occupied by holes). Diagrams (c) and (d) show four other absorption processes available at non-equilibrium population after the UV-radiation (arrows $E_{Vis}$ and $E_{ind}$), when defects D1 and D2 are in the charge states $i+1$ and $j-1$. Here, D1 and D2 can be excited thermally or optically (arrows $E_{Vis}$) to initiate the thermal or optical restoration via CB-mediated (c) and VB-mediated (d) reverse charge transfer. Emission $E_{res}$ with energies higher than the absorbed energy $E_{Vis}$ is possible during the restoration.

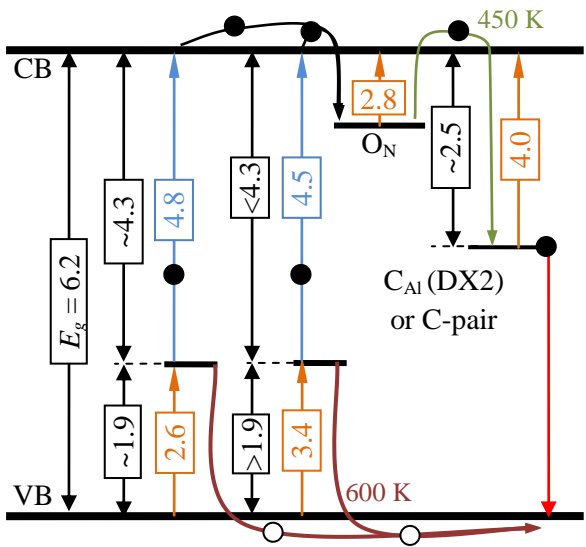

Fig. 8. The diagram of observed absorption peaks in area B2. The double-headed arrows illustrate the position of defect states ($\Delta E$). Single-headed arrows show the absorption energies from the initial states (blue) or after UV-irradiation (orange). The transport of electrons and holes is shown by the lateral arrow paths within CB and VB.

TABLES

Table I. SIMS concentrations of impurtites: carbon [C], oxygen [O], and silicon [Si], as well as R = ([Si]+[O])/[C]) for three sample areas B1 – B3.

| Area | [C], $10^{17}$cm$^{-3}$ | [O], $10^{17}$cm$^{-3}$ | [Si], $10^{17}$cm$^{-3}$ | R |
|------|------|------|------|------|
| B1 | 57 | 21 | 0.7 | 0.4 |
| B2 | 18 | 33 | 3.7 | 2.0 |
| B3 | 18 | 64 | 3.3 | 3.7 |

Table II. Absorption peak positions and spectral ranges as observed in areas B1, B2, B3 of the AlN crystal, probable optical transition type (see Figs. 7 and 8), existence of the band and variation of its intensity before and after UV radiation, respectively, DOP, and effective annealing temperature T for complete restoration.

*The UV-induced absorption change is negligible and restores at room temperature.

**from the Gaussian Fitting in Fig. 5 (a)

| Peak, eV | Type | Area | Before UV or after restoration | After UV | Spectral range, eV | DOP | T, K |
|---|---|---|---|---|---|---|---|
| 2.6 | $E_{Vis}(D1)$ | B1 | Yes* | Unchanged* | 2.1–3.0 | ~ 0.1 | |
| | | B2 | No | Appeared | 2.1–3.0 | ~ 0.1 | 500-600 |
| | | B3 | No | Not present | - | - | - |
| 2.8 | $E_{Vis}(D2)$ | B1 | No* | Unchanged* | 2.2–3.2 | - | 300 |
| | | B2 | No | Appeared | 2.2–3.2 | ~ 0* | 400-450 |
| | | B3 | Weak | Amplified | 2.2–3.2 | ~ -0.10 | 400-450 |
| 3.4 | $E_{Vis}(D1)$ | B1 | Yes* | Unchanged* | 3.0–3.6 | ~ 0.35 | - |
| | | B2 | No | Appeared | 3.0–3.6 | ~ 0.35 | 450-600 |
| | | B3 | - | - | - | - | - |
| 3.2 | $E_{Vis}(D2)$ | B3 | Yes | Amplified | 2.0–4.5 | ~ 0.30 | - |
| 4.0 | $E_{Vis}(D2)$ | B2 | No | Amplified | 3.8–4.2 | < 0.30 | 600 |
| 4.5** | $E_{UV}(D1)$ | B1 | > 1200 cm$^{-1}$ | Too intense | 4.1–4.8** | - | - |
| | | B2 | ~ 120 cm$^{-1}$ | Attenuated | 4.1–4.8** | > 0.50 | 500-600 |
| | | B3 | No | Not present | - | - | - |
| 4.8** | $E_{UV}(D1)$ | B1 | >1200 cm$^{-1}$ | Too intense | 4.5–5.1** | - | |
| | | B2 | ~ 100 cm$^{-1}$ | Attenuated | 4.5–5.1** | ~ 0.0 | 500-600 |
| | | B3 | No | Not present | - | - | - |
| 5.1 | $E_{ind}(D1)$ | B2 | - | Amplified | 5.0–5.4 | < 0.2 | 450-600 |